\documentclass[twocolumn]{aastex631}
\usepackage{longtable}
\usepackage{amsmath}    
\usepackage{commath}
\usepackage{soul}
\usepackage{CJK}
\usepackage{makecell}
\usepackage{multirow}
\usepackage{txfonts}
\usepackage{chngcntr}

\newcommand{\Ha}{\makebox{H$\alpha\,$}}
\newcommand{\Hb}{\makebox{H$\beta\,$}}

\def\OIII{[O\,{\sc iii}]}
\def\OI{[O\,{\sc i}]}
\def\Ha{H{$\rm{\alpha}$}}
\def\Hb{H{$\rm{\beta}$}}

\def\NII{[N\,{\sc ii}]}
\def\SII{[S\,{\sc ii}]}
\def\HeII{He\,{\sc ii}$\lambda$4686}

\begin{document}
\begin{CJK*}{UTF8}{gbsn}
\title{Extended Emission Line Region in a Post-Starburst Galaxy Hosting Tidal Disruption Event AT2019qiz and Quasi-periodic Eruptions}
\author[0000-0002-0055-7277]{Yifei Xiong (熊翊飞)}
\affiliation{Key Laboratory for Research in Galaxies and Cosmology, Shanghai Astronomical Observatory, Chinese Academy of Sciences, 80 Nandan Road, Shanghai 200030, People's Republic of China}
\affiliation{University of Chinese Academy of Sciences, 19A Yuquan Road, Beijing 100049, People's Republic of China}
\author[0000-0002-7152-3621]{Ning Jiang (蒋凝)}
\affiliation{Department of Astronomy, University of Science and Technology of China, Hefei, 230026, People's Republic of China}
\affiliation{School of Astronomy and Space Sciences, University of Science and Technology of China, Hefei, 230026, People's Republic of China}
\author[0000-0001-9608-009X]{Zhen Pan (潘震)}
\affiliation{Tsung-Dao Lee Institute, Shanghai Jiao-Tong University, Shanghai, 520 Shengrong Road, 201210, People's Republic of China}
\affiliation{School of Physics \& Astronomy, Shanghai Jiao-Tong University, Shanghai, 800 Dongchuan Road, 200240, People's Republic of China}
\affiliation{Tsung-Dao Lee Institute, School of Physics and Astronomy, Shanghai Jiao Tong University,
Key Laboratory for Particle Physics, Astrophysics and Cosmology (Ministry of Education),
Shanghai Key Laboratory for Particle Physics and Cosmology,
800 Dongchuan Road, Shanghai, 200240, P.R. China}
\author[0000-0003-2478-9723]{Lei Hao (郝蕾)}
\affiliation{Key Laboratory for Research in Galaxies and Cosmology, Shanghai Astronomical Observatory, Chinese Academy of Sciences, 80 Nandan Road, Shanghai 200030, People's Republic of China}
\author[0000-0002-8237-0237]{Zhenzhen Li (李臻臻)}
\affiliation{Key Laboratory for Research in Galaxies and Cosmology, Shanghai Astronomical Observatory, Chinese Academy of Sciences, 80 Nandan Road, Shanghai 200030, People's Republic of China}

\begin{abstract}
We present a comprehensive analysis of the extended emission line region (EELR) in the host galaxy of the tidal disruption event (TDE) AT2019qiz, utilizing Very Large Telescope/MUSE integral-field spectroscopy. The high spatial-resolution data reveal a bi-conical emission structure approximately $3.7~\mathrm{kpc}$ in scale within the galactic center, characterized by a prominent \OIII\ line in the nucleus and significant \NII\ line emission extending into the EELR. Spectral analysis of the EELR indicates line ratios consistent with Seyfert ionization in the center and LINER-type ionization in the outer diffuse region, suggesting ionization from galactic nuclear activity. The required ionizing luminosity, estimated from the \Ha\ and \Hb\ luminosities based on the photoionization and recombination balance assumption, is $10^{41.8}$ $\mathrm{erg\,s^{-1}}$ for all spaxels classified as active galactic nucleus (AGN), and $10^{40.7}$ $\mathrm{erg\,s^{-1}}$ for spaxels in the central $0.9~\mathrm{kpc}$ Seyfert region. However, the current bolometric luminosity of the nucleus $L_{\text{bol}} \leq  10^{40.8}\,\mathrm{erg\,s^{-1}}$, estimated from quiescent-state soft X-ray observations, is insufficient to ionize the entire EELR, implying a recently faded AGN or a delayed response to historical activity. Stellar population analysis reveals a post-starburst characteristic in the EELR, and the gas kinematics show disturbances and non-circular components compared to the stellar kinematics. Notably, the recent detection of quasi-periodic eruptions (QPEs) in the X-ray light curve of AT2019qiz confirms the TDE-QPE association. Our findings provide direct evidence for an AGN-like EELR in the host galaxy of the nearest TDE with QPE detection, offering new insights into the complex interplay between TDEs, QPEs, AGN activity, and host galaxy evolution.
\end{abstract}

\section{INTRODUCTION} \label{sec:sec1}

\correspondingauthor{Ning Jiang, Yifei Xiong} \email{jnac@ustc.edu.cn, xiongyf@shao.ac.cn}

Tidal disruption events (TDEs) and quasi-periodic eruptions (QPEs) are two distinct transient phenomena observed in galactic nuclei in recent years. TDEs are well known phenomena which occur when stars are tidally disrupted by supermassive black holes (SMBHs) \citep{Rees1988,Gezari2021}, producing characteristic light curves across multiple wavelengths. QPEs, on the other hand, are characterized by nearly periodic X-ray eruptions with recurrence times of hours to days, whose physical origin remains debated. Current theoretical explanations for QPEs primarily include instabilities originating from the accretion disk itself \citep[e.g.][]{Sniegowska2020,Raj2021,Pan2022,Pan2023}, mass transfer from a partially disrupted star or white dwarf via Roche-lobe overflow \citep[e.g.][]{King2022,Krolik2022,Wang2022,Metzger2022,Linial2023}, and periodic impacts of a single stellar mass object with the accretion disk of the SMBH, such as a main sequence star or a stellar mass black hole \citep[e.g.][]{Lu2023,Linial&Metzger2023,Franchini2023,Tagawa2023,Zhou2024a,Zhou2024b,Zhou2025}. All theoretical models require the presence of an accretion disk, which can form either through the direct accretion of gas by the SMBHs or as a consequence of the tidal disruption of stars during a TDE. Recent observations suggest a potential connection between TDEs and QPEs. For instance, TDE-like declining light curves have been observed in QPE events (GSN069; \citealt{Shu2018,Miniutti2023}, eRO-QPE3; \citealt{Arcodia2024}), and QPE-like eruptions have been detected in two TDEs (XMMSL1 J024916.6-04124; \citealt{Chakraborty2021}, AT2019vcb; \citealt{Quintin2023}). The most important progress recently was that \citet{Nicholl2024} observed nine quasi-periodic X-ray eruptions from 29 February 2024 to 14 March 2024 in the host galaxy of TDE AT2019qiz, which is one of the nearest TDE discovered on 2019-9-19 UT and reach its peak luminosity at 2019-10-08 UT \citep{Nicholl2020}. This is the first confirmed repeating QPE in a spectroscopically confirmed TDE. After that, repeated QPEs after TDE have also been reported in AT2022upj \citep{Chakraborty2025}. 

Recent studies about the host galaxy properties of the TDEs \citep[e.g.][]{French2020a} and the QPEs \citep[e.g.][]{Wevers2022} have revealed intriguing similarities between them \citep{Wevers2024b}. Both types of events predominantly occur in post-starburst or E+A galaxies \citep[e.g.][]{French2016, French2017} characterized by high stellar surface mass densities, $\text {S\'ersic}$ indices, and high bulge-to-total light (B/T) ratios \citep{Law-Smith2017, Graur2018, Gilbert2024}. Intriguingly, extended emission line regions (EELRs) \citep[e.g.][]{Lintott2009,Keel2012,Keel2015}, also named as extended narrow line regions (ENLRs), consisting of ionized gas spanning scales of several to ten kpc, have been detected in several host galaxies of both TDEs and QPEs. To date, three TDEs (ASASSN–14li; \citet{Prieto2016}, AT2019azh; \citet{French2023}, and iPTF–16fnl; \citet{Wevers2024a}) have been reported to exhibit EELRs in their host post-starburst galaxy. Among the about 12 known QPEs and candidates \citep{Sun2013,Miniutti2019, Giustini2020, Arcodia2021, Arcodia2022, Arcodia2024, Chakraborty2021, Evans2023, Guolo2024, Nicholl2024,Hernandez-Garcia2025,Chakraborty2025}, three of them (GSN 069; \citet{Patra2024}, RXJ1301, and eRO-QPE2; \citet{Wevers2024b}) have been detected with EELRs using Hubble Space Telescope (HST) narrow-band images and the MUSE integral-field spectrograph. The fractional incidence of EELRs in QPE and TDE host galaxies appears higher than in the general galaxy population \citep{Wevers2024a,Wevers2024b}, despite the limited sample size and the fact that MUSE observations provide deeper detection thresholds than narrowband-imaging surveys \citep[e.g.][]{Keel2024}, making a comparison less direct.

All the EELRs in QPE and TDE host galaxies require ionization from a non-stellar continuum source, such as radiation from active galactic nucleus (AGN) accretion disks. However, the nuclear luminosities currently observed in these sources are too low to maintain the necessary ionization \citep{Wevers2024a,Wevers2024b}, pointing toward a recently faded AGN scenario \citep{Keel2017,French2023}. Furthermore, some theoretical models require the presence of an accretion disk prior to the onset of  QPEs. For example, the leading EMRI+disk model \citep{Linial&Metzger2023,Franchini2023} suggests that an extreme mass ratio inspiral (EMRI), in which a stellar-mass object orbits a SMBH, collides with an accretion disk formed from a TDE twice per orbit. Detailed orbital analyses show that a large fraction of QPE EMRIs exhibit low orbital eccentricity, supporting the hypothesis that they were born in or were captured by a past AGN disk \citep{Zhou2024a,Zhou2024b,Zhou2025}, named wet EMRI formation channel \citep{Sigl2007,Levin2007,Pan2021a,Pan2021b}. In short, these findings suggest a potential physical connection among host galaxy properties, past nuclear activity, TDEs, and the occurrence of QPEs. The burst of star formation together with the galactic nuclear activities and the elevated TDE rate produces UV photons from accretion disks, leading to the ionization of surrounding gas. 

Therefore, EELRs can serve as indirect probes of ongoing or past nuclear activity. However, no EELR has been previously reported in AT2019qiz, the first unambiguous QPE found in a standard spectroscopically confirmed TDE. Unlike previous sources where either the QPE detection was less definitive (e.g., AT2019vcb) or the TDE classification was not as robust (e.g., GSN069), AT2019qiz represents an ideal target to probe potential fading AGN signatures and investigate the physical connections between TDEs, QPEs, and AGN activity in a single well-characterized system. In this work, we present a detailed study of the AT2019qiz host galaxy (LEDA 980815) using integral-field spectroscopy data from MUSE. The face-on and barred galaxy is located at $\mathrm{RA} = 04^{\mathrm{h}} 46^{\mathrm{m}} 37.880^{\mathrm{s}}, \mathrm{DEC} = -10^{\circ} 13^{\prime} 34.92^{\prime\prime} \, (\mathrm{J}2000)$, with a redshift of $\textit{z} = 0.01513$. Previous studies have shown that it harbors a supermassive black hole of approximately $10^6\,\rm{M_{\odot}}$ \citep{Nicholl2020,Hung2021,Short2023}, and mid-infrared echoes \citep{Jiang2016,Jiang2021} in the WISE light curve suggest the presence of a dusty torus characteristic of AGN \citep{Short2023,Pasham2025}. Using archival MUSE data, we demonstrate the existence of an EELR in the reconstructed images in Section \ref{sec:sec2}, where we also introduce our detailed spectral fitting and analysis methodology. In Section~\ref{sec:sec3}, we present our spectral analysis results, including BPT classification of selected regions and the entire datacube, stellar population analysis, kinematics, and an estimation of the ionizing luminosity assuming AGN activity as the ionization source. We adopt a cosmology with $H_{\rm 0} \rm{=70~km~s^{-1}~{Mpc}^{-1},~\Omega_m=0.3,~and~\Omega_\Lambda=0.7}$, corresponding to a luminosity distance of 65.55 Mpc and an angular diameter distance of 63.61 Mpc, which we use to calculate luminosities and physical sizes of the EELR throughout our analysis.

\section{MUSE Data and Analysis} \label{sec:sec2}

\begin{figure*}[ht!]
\begin{center}
\epsscale{1.0}
\plotone{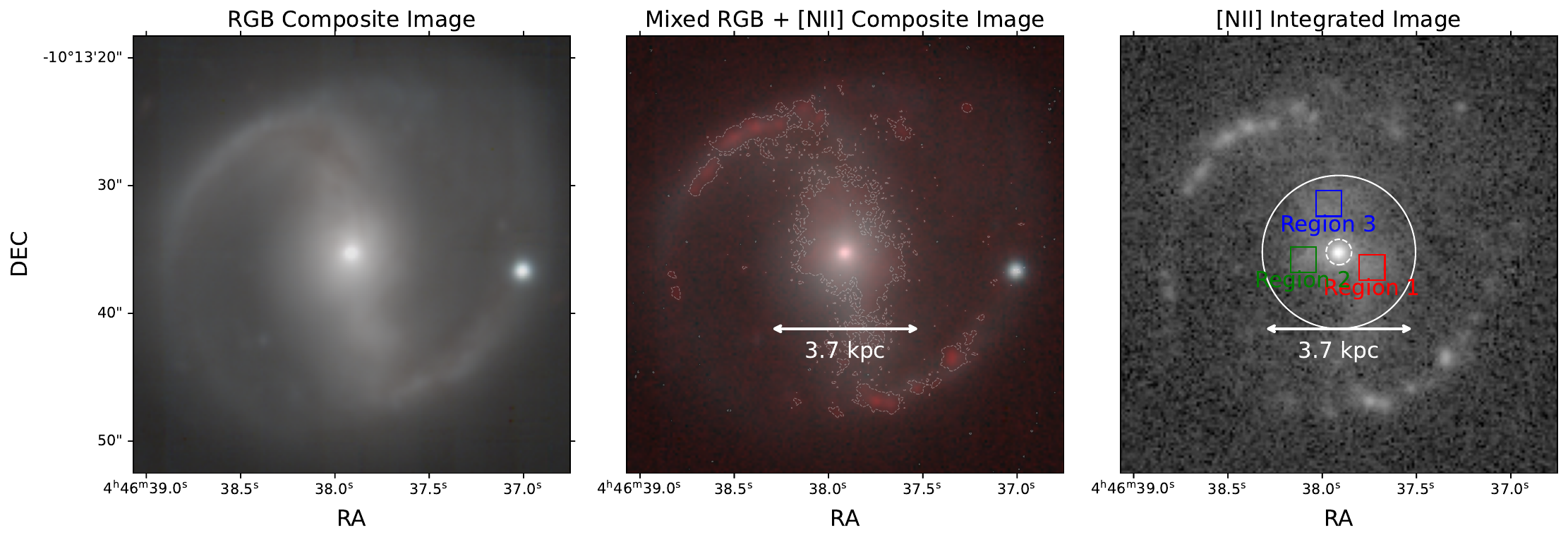}
\end{center}
\caption{Left: Red (Bessel R), Green (Bessel V), and Blue (Bessel B) composite image reconstructed from the MUSE data cube of the host galaxy of AT2019qiz, showing a barred disk galaxy with two spiral arms. Middle: RGB composite image overlaid with \NII\ integrated flux map in red; contours show the log$_{10}$(\NII) flux distribution, revealing a $\sim$ 3.7 kpc bi-conical \NII\ structure in the center that indicates the presence of an EELR in this galaxy. Right: \NII\ integrated flux map. Extended \NII\ emission is distributed around the galaxy center (within the white circle). Regions marked in the right panel are used for spectral analysis, where the red, green, and blue boxes are 11 $\times$ 11 spaxels in size. The region inside the white solid circle denotes the entire EELR, while spectra inside the dashed circle contain broad line components originating from the TDE. In Figure \ref{fig:figA1}, we also present \OIII\ and \Ha\ integrated flux maps. All images are displayed using an arcsinh stretch.}
\label{fig:fig1}
\end{figure*}

We obtained the integral-field spectrograph VLT MUSE \citep{Bacon2010} Phase3 datacube from the ESO archive, which was observed on 2021-02-08 (Program ID:105.20GS, PI: Sam Kim), 481 rest-frame days after the TDE reach its peak luminosity, with an effective exposure time of 2999 s. The spectral coverage ranges from 470 to 935 nm with a spectral resolution of R = 3014. The observation was conducted in WFM-AO-N mode, and the sodium light contaminated region between 582 and 597 nm was excluded. The spatial resolution is $0.2^{\prime\prime}$ $\textrm{spaxel}^{-1}$, and the DIMM seeing is $0.6^{\prime\prime}$. We performed multiple analyses on the datacube. First, we generated emission line maps by integrating the continuum-subtracted spectra around each emission line, as shown in Section~\ref{subsec:sec2_1}. We then conducted detailed spectral fitting on both individual regions of interest and the entire datacube after 3×3 spatial binning to improve the signal-to-noise ratio (SNR), which will be presented in Section~\ref{subsec:sec2_2}.

\subsection{Datacube Quicklook: An Extended Emission Line Region}\label{subsec:sec2_1}
We first performed a quick inspection of the datacube. For each emission line, we obtained line flux distribution maps by integrating the continuum-subtracted emission line profiles, where the continuum was estimated through linear interpolation of the adjacent spectral regions. The \OIII $\lambda$5007, \Ha\ and \NII $\lambda$6584 maps are presented in the Appendix Figure \ref{fig:figA1}. This straightforward method helps avoid potential artifacts from poor Gaussian line fitting. The maps reveal a prominent EELR at the center of the host galaxy, particularly evident in the \NII\ emission. In Figure~\ref{fig:fig1}, we cropped the datacube of the galaxy region, and made a composite RGB color image to illustrate the distribution of the EELR relative to the host galaxy. This color image was created by convolving the datacube with Bessel R, V, and B filter transmission curves from the SVO Filter Profile Service \citep{Rodrigo2020}. The integrated \NII\ flux map is overlaid on the color image of the galaxy. The EELR appears to be distributed within the galactic bar at the center, exhibiting a bi-conical structure extending approximately $\sim 3.7~\mathrm{kpc}$, the boundary is estimated just by visual inspection.

\subsection{Detailed Spectral Fitting and Datacube Analysis} \label{subsec:sec2_2}

\begin{figure*}[ht!]
\begin{center}
\epsscale{1}
\includegraphics[width=0.9\textwidth]{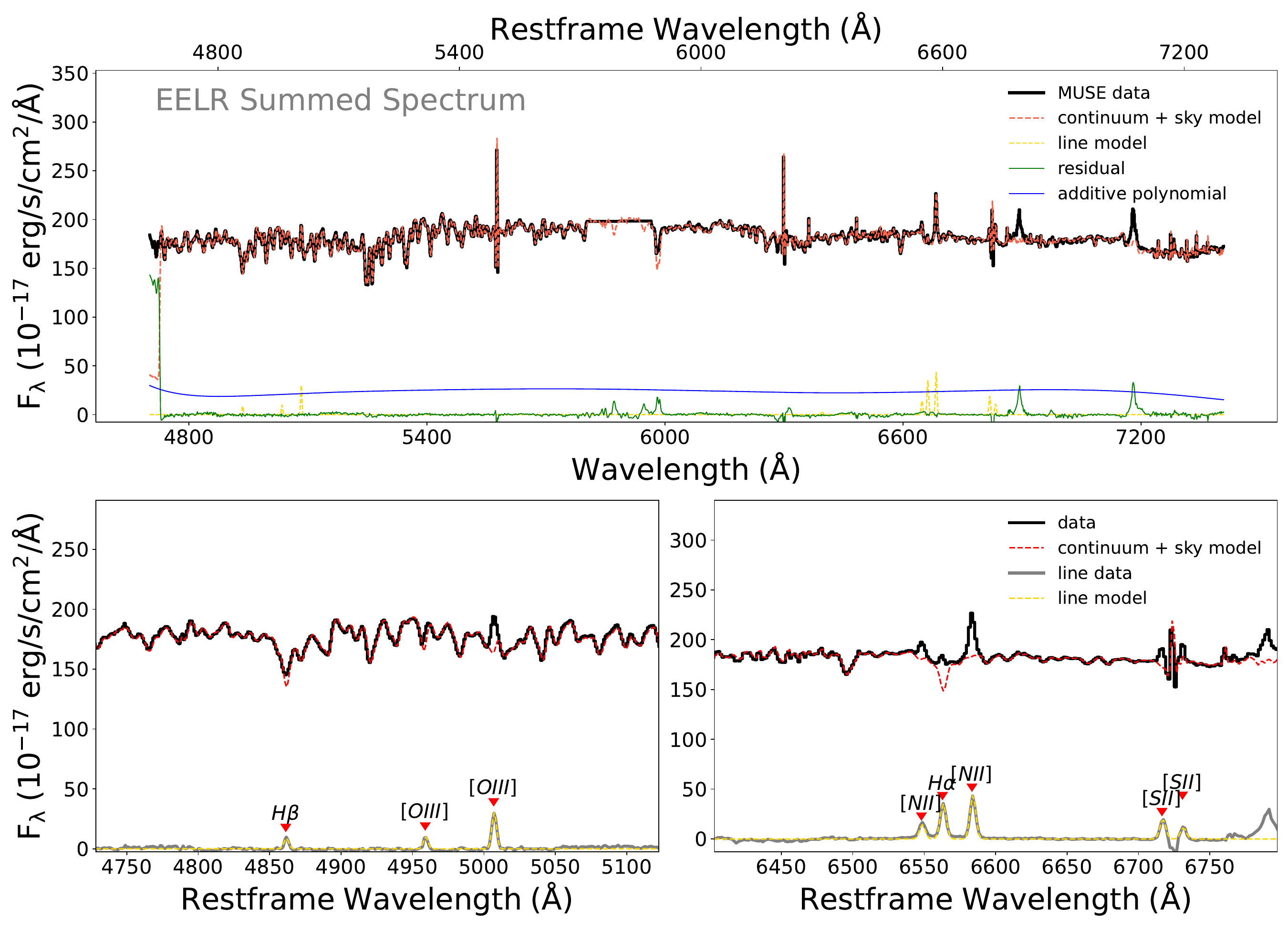}
\end{center}
\caption{The pPXF spectral fitting result of the entire EELR (named ``EELR summed''), except for the center broad line region. The black solid line represents the MUSE spectral data, the red dashed line shows the continuum+sky model, the gray solid line indicates the emission line data (obtained by subtracting the continuum+sky model from the data), and an 8-order additive polynomial is shown in blue solid line. The yellow dashed line represents the emission line model, and the green line displays the fitting residuals. Fitting results for Region 1,2,3 and center are shown in Appendix Figure \ref{fig:figA2}.}
\label{fig:fig2}
\end{figure*}

We performed a visual inspection of the spectra inside the datacube, there is broad Balmer emission line components in the galaxy center, originating from a broad line region formed after the TDE AT2019qiz \citep{Nicholl2020,Short2023}. The rest of the spectra only exists narrow lines. We choose some region-of-interests to perform further analysis. As shown in Figure \ref{fig:fig1}, Region 1 (red box), Region 2 (green box), and Region 3 (blue box) were selected to assess local properties of the EELR, all the spectral inside each region box (with 11 $\times$ 11 spaxels) were summed together, to improve the signal-to-noise ratio of the spectra. Specifically, we combined all spectra between the solid and dashed white circles in Figure \ref{fig:fig1} to characterize the overall properties of the EELR, and region within the dashed circle contains broad line components. Our subsequent spectral analysis followed two approaches: spectral fitting of selected regions and fitting of the entire datacube. 

Prior to fitting, all spectra were corrected for Galactic extinction \citep[$E(B-V)=0.11$;][]{Fitzpatrick1999, Schlafly2011}. We utilized the pPXF code \citep{Cappellari2017,Cappellari2023} to perform simultaneous fitting of the stellar population continuum and emission lines across the spectra range from 4700\,\AA\ to 7400\,\AA. The continuum spectra were fitted using E-MILES stellar library \citep{Vazdekis2010,Vazdekis2015,Vazdekis2016} and multiplied by the attenuation curve of \citet{Calzetti2000} to account for the dust attenuation of the host galaxy. Sky line contamination was evident in the spectra, so we constructed a sky spectrum template from the sky background of the datacube and included an 8th-order additive polynomial in the fitting. Narrow emission lines (\Ha, \Hb, \NII, \SII, \OIII) were modeled using line templates with an average instrument FWHM of 2.65\,\AA, convolved with a Gaussian line-of-sight velocity distribution (LOSVD) function. For spectra within the white dashed circle in Figure \ref{fig:fig1}, two broad Balmer line components for each \Ha\ and \Hb\ lines were included in the fitting. Through this fitting procedure, we obtained measurements of emission line fluxes, stellar population properties including age and metallicity, as well as the kinematics of both gas and stars. 

The spectra for each region and the fitting results are presented in Figure~\ref{fig:fig2} for the entire EELR and Figure~\ref{fig:figA2} for Region 1,2,3 and center. It should be noted that the narrow \Hb\ lines cannot be seen in the original spectral data, as there are only absorption lines. However, they can be revealed after subtracting the modeled stellar continuum. We believe this may overestimate the flux of the \Hb\ lines, so the \OIII/\Hb\ ratio in the BPT diagram may be underestimated. Similarly, the \Ha\ emission lines exhibit lower flux compared to \NII\ in the original data. After correcting for the absorption feature, the residual \Ha\ emission will show enhanced flux levels than before. The \HeII\ line, which is also used as an AGN indicator \citep[e.g.][]{Shirazi2012,Tozzi2023} and can be quite strong in some EELRs \citep{Keel2012}, lies near the edge of our wavelength range when redshifted (4756\,\AA). We detect only a broad \HeII\ component in the galactic centre, which  interpreted as an outflow from AT2019qiz \citep{Nicholl2020}, while the EELR shows no significant \HeII\ emission. The position of the line close to the wavelength limit makes our stellar population continuum fitting poor, and also prevents an accurate measurement of the \HeII\ flux. Therefore, we omit the analysis of \HeII\ in this paper. We also performed measurements on the entire datacube. To improve fitting efficiency, we cropped the sky background region without useful signal, focusing only on the region containing the galaxy, and applied 3 $\times$ 3 spaxel binning to the original datacube to enhance the signal-to-noise ratio, which degrade the spatial resolution to $0.6^{\prime\prime}$ $\textrm{spaxel}^{-1}$. We then conducted batch measurements using pPXF with the same settings as mentioned above. The broad line components in the central spaxels were also considered.

\section{RESULTS and DISCUSSION} \label{sec:sec3}

\subsection{BPT Diagrams and Stellar Population of the EELR} \label{subsec:sec3_1}

\begin{figure*}[ht!]
\begin{center}
\epsscale{1.0}
\includegraphics[width=1.0\textwidth]{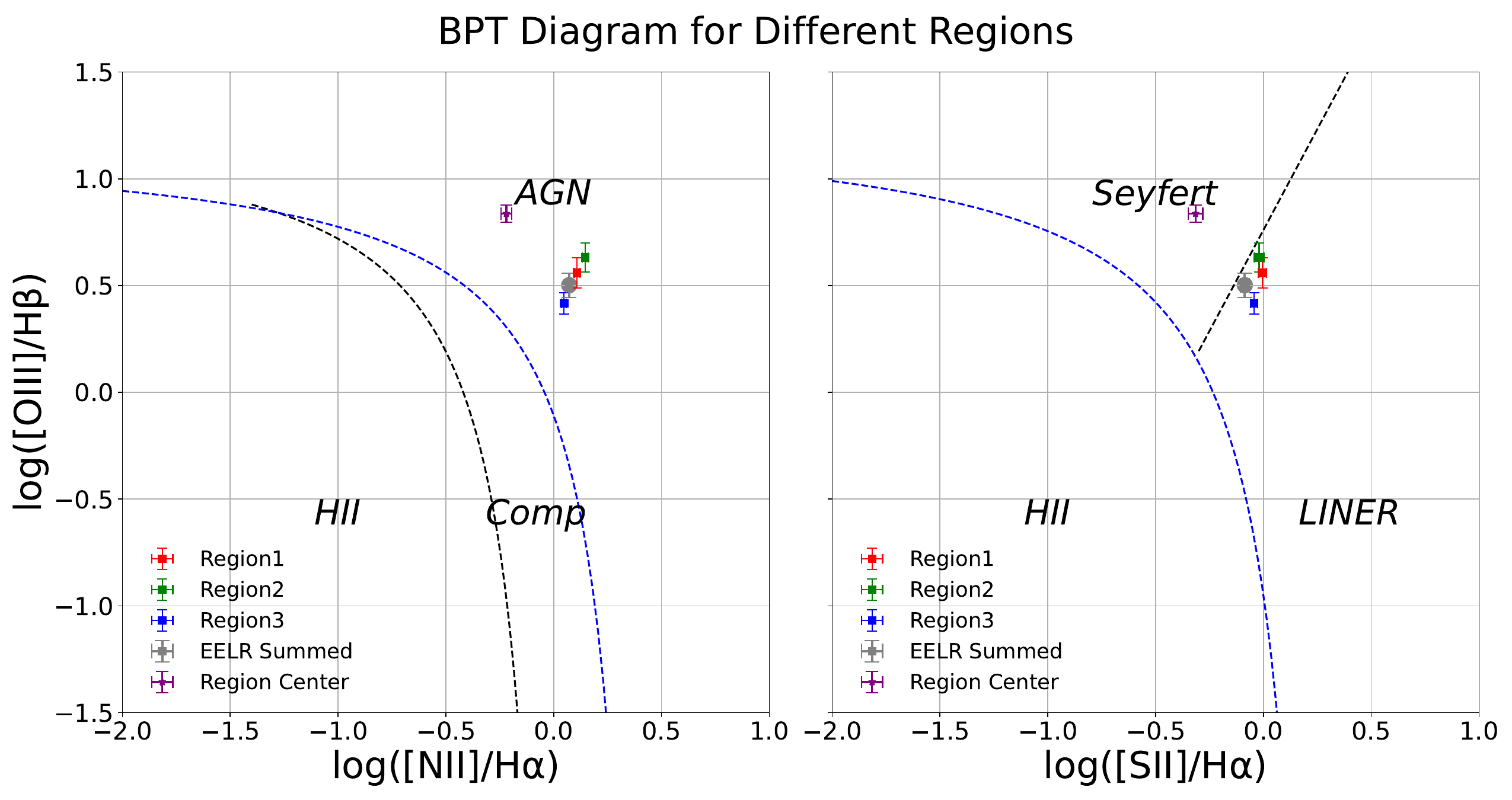}
\end{center}
\caption{BPT diagram showing line ratios for all five regions. Region 1, 2, and 3 are represented by red, green, and blue points respectively, while the entire EELR region is shown in gray points, and the central region with broad line (Region Center) is displayed in purple. The summed spectra of the entire EELR, along with the local spectra from Regions 1, 2, and 3, are classified as AGN in the \NII/\Ha\ diagram and LINER in the \SII/\Ha\ diagram. However, the narrow emission lines in the galaxy center are classified as Seyfert. The classification lines in the diagrams are from \citet{Kewley2001,Kewley2006,Kauffmann2003}.}
\label{fig:fig3}
\end{figure*}

Based on our spectral measurements, we obtained narrow emission line fluxes from the entire EELR (regions between the solid and dashed circles), as well as from Regions 1, 2, 3, and within the dashed circle (Region center with broad line). We calculated the line ratios of \OIII/\Hb, \SII/\Ha, and \NII/\Ha, and plotted them on BPT diagnostic diagrams \citep{Baldwin1981} as shown in Figure~\ref{fig:fig3}. Classification lines defined in \citet{Kewley2001,Kewley2006,Kauffmann2003} were used to classify the spectra. For the entire EELR region, as well as local Regions 1, 2, and 3 inside it, the line ratios fall within the AGN region in \NII/\Ha\ diagrams and lie near the boundary between Seyfert and LINER classifications in \SII/\Ha\ diagrams. However, the narrow emission line ratios from the central broad line region exhibit clear Seyfert-like ionization characteristics. As mentioned in Section \ref{subsec:sec2_2}, the \OIII/\Hb\ ratio may be underestimated. Therefore, this result supports the possibility that the EELR region is ionized by AGN activity. \citet{Nicholl2020} used long-slit X-shooter spectra and found the line ratio of this galaxy located in the region between star-forming galaxies and AGN, indicating a weak AGN signal. The spectra from the long-slit spectrograph could be contaminated by star-forming regions, whereas the spatially resolved MUSE data can avoid this and reveal the EELR region dominated by AGN classification. \citet{Hung2021} and \citet{Short2023} also measured the spectrum at the galaxy nucleus and found the line ratio located in the Seyfert region of the BPT diagram, consistent with our results here. We will further demonstrate the spatially resolved spectroscopic classification in the following section. 

Additionally, the stellar population analysis reveals a star formation event approximately 1 Gyr ago, as indicated by the fitting results from the entire EELR region, as well as local Regions 1, 2, 3, and the central region. Figure \ref{fig:figA3} illustrates the stellar population result for the entire EELR region. The pPXF fitting enables us to derive weights for various E-MILES stellar population spectral templates \citep{Vazdekis2016}. Consequently, we can plot weight maps for ages and metallicities, and display weights for different ages, which can be interpreted as the star formation history. It is evident that after the initial star formation 10 Gyr ago, another star formation event occurred 1 Gyr ago, after which star formation was quenched. The different weights of the stellar spectra correspond to different mass-to-light ratios. Through this analysis, we estimate that this star formation event approximately 1 Gyr ago formed stars accounting for about 10\% of the galaxy's total stellar mass. This pattern is characteristic of post-starburst galaxies \citep{Bergvall2016}. We also found that the host of AT2019qiz is located in the ``green valley'' region, approaching the ``red sequence'' in the galaxy color-magnitude diagram \citep{Bell2004}. This classification is based on g and r band data (after corrections for Galactic extinction) from the DESI Legacy Survey \citep{Dey2019}, which is consistent with the classification in \citet{Hammerstein2021}. As mentioned in the introduction, TDEs and QPEs are more likely to occur in post-starburst galaxies, a finding corroborated by our analysis. However, the spectral wavelength range of the MUSE spectrum does not include $H\delta_{\text{A}}$ or $D_n(4000)$ for our galaxy. Future spectral observations with broader wavelength coverage are necessary for more reliable stellar population synthesis.

\subsection{Emission Line Flux and BPT Classification Map} \label{subsec:sec3_2}

\begin{figure*}[ht!]
\begin{center}
\epsscale{1.0}
\plotone{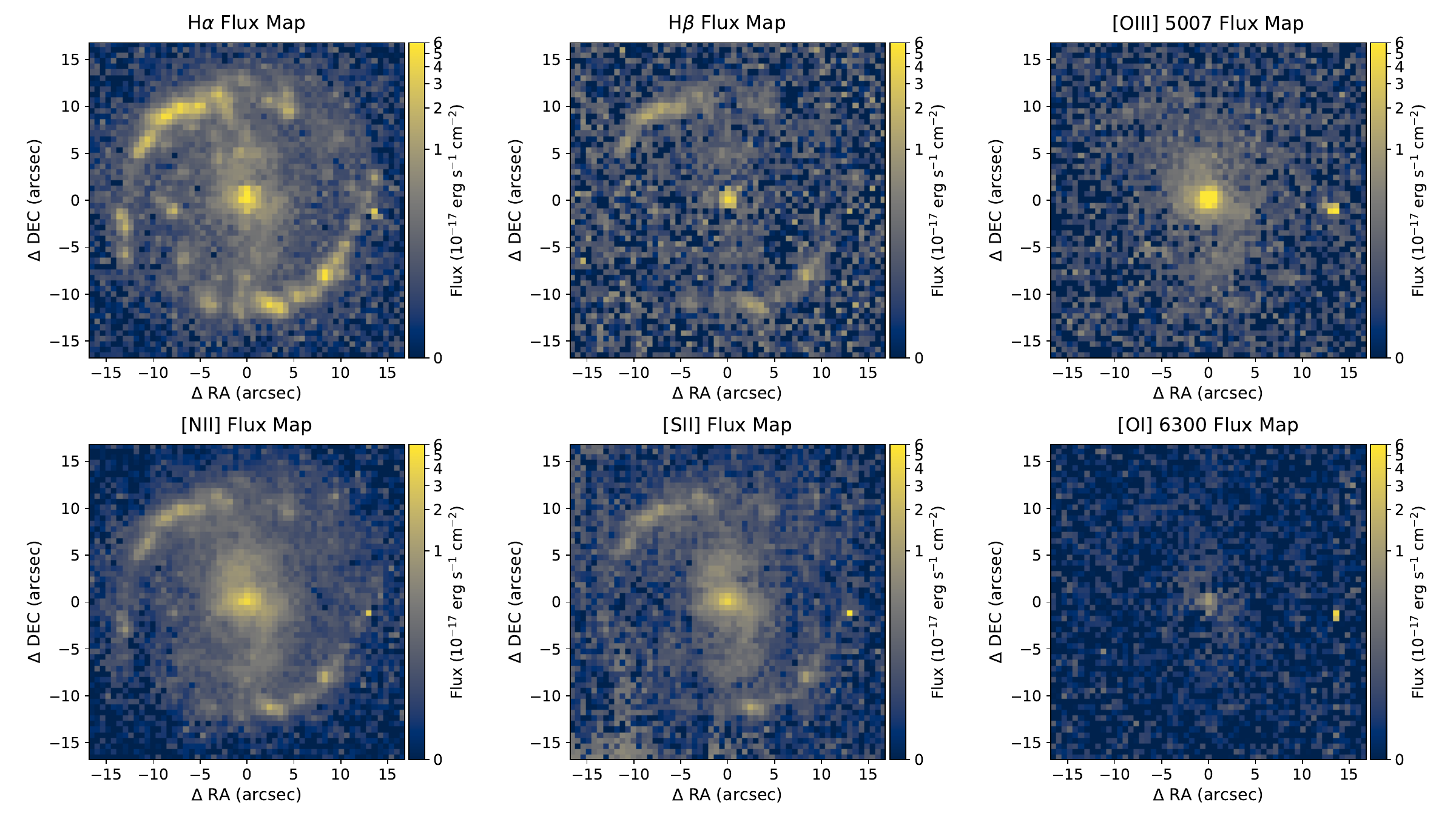}
\plotone{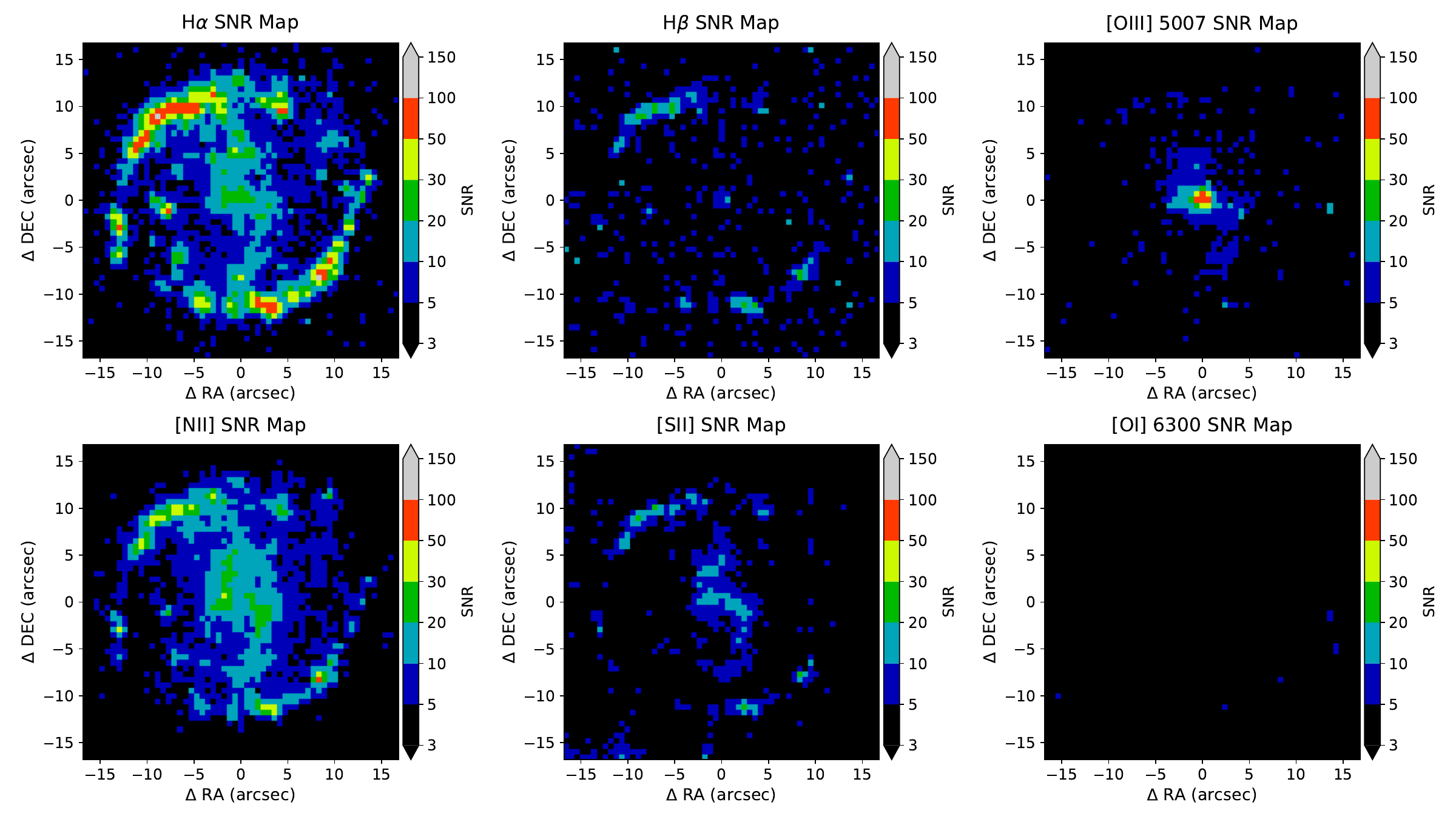}
\end{center}
\caption{Narrow emission line flux(arcsinh streched) and signal-to-noise ratio(SNR) maps of \Ha, \Hb, \OIII, \NII, \SII, and \OI\ derived from spectral fitting of the 3$\times$3 binned datacube after removing the broad line components in the central region. All flux maps reveal an EELR with a size greater than $3~\mathrm{kpc}$, with a compact core in the center of the galaxy. HII region clumps are distributed along the spiral arms of the host galaxy. All map are arcsinh streched. In the EELR, \OIII\ emission is strong in the center, while diffuse \NII\ emission is obvious in the outer regions. The SNR maps displayed that \Ha, \NII\ and \SII\ fluxes in the EELR have similar distribution, whereas the \OIII\ flux is enhanced in the center of the galaxy. The extended \OIII\ structure is only evident around the galactic center and does not extend to the spiral arms.}
\label{fig:fig4}
\end{figure*}

\begin{figure*}[ht!]
\begin{center}
\epsscale{1.0}
\plotone{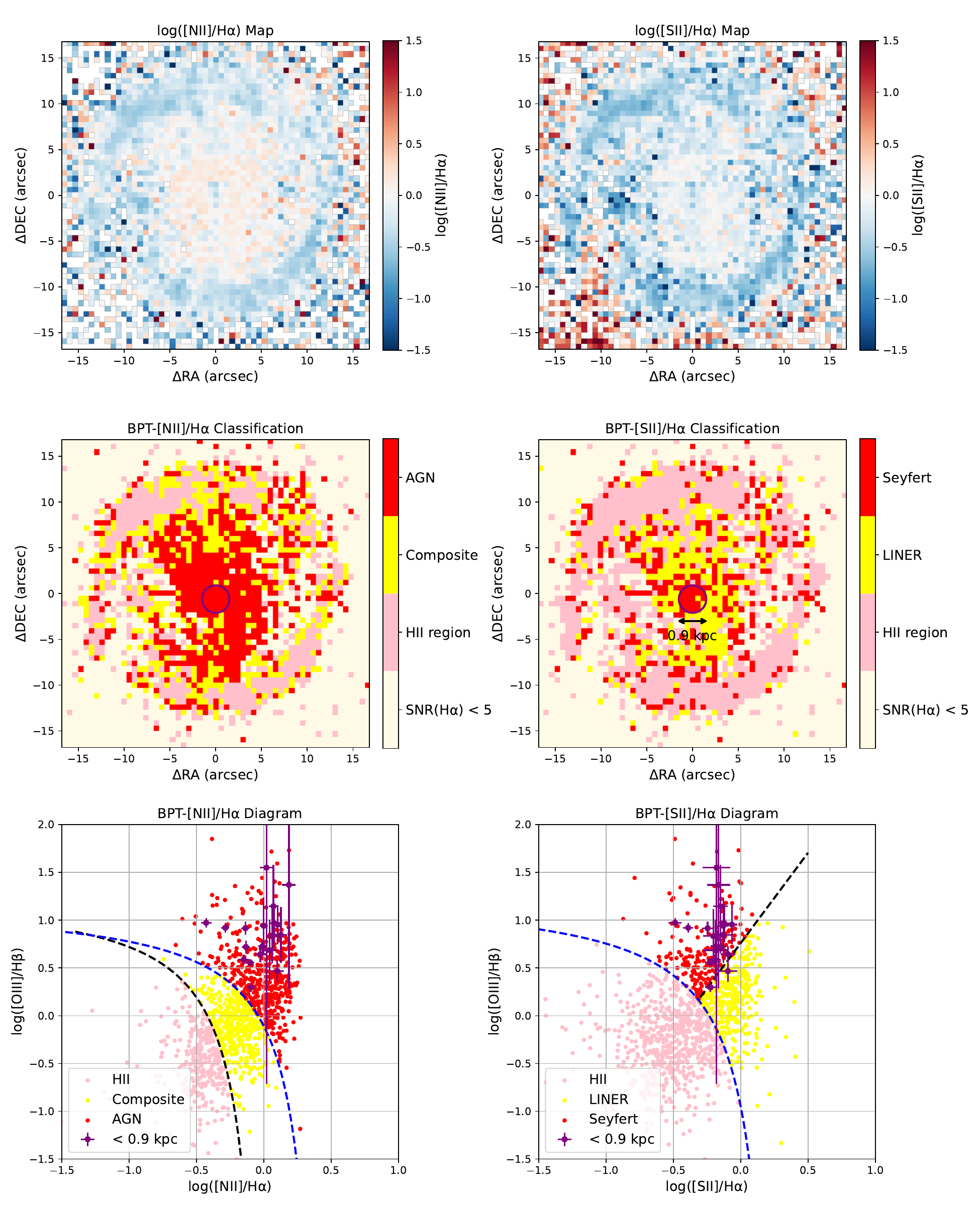}
\end{center}
\caption{The top panels show the spatial distribution of emission line ratios log$_{10}$(\NII/\Ha) and log$_{10}$(\SII/\Ha). The middle panels display the BPT classification map for individual spaxels with SNR(\Ha) $>$ 5, where different colors represent different ionization mechanisms. The bottom panels show the BPT diagram for each individual spaxel. For the \NII/\Ha\ classification, the EELR is dominated by AGN ionization (red), while the spiral arms are dominated by star formation activities (pink). For the \SII/\Ha\ classification, the EELR can be separated into two parts, with the central $\sim 0.9~\mathrm{kpc}$ inside the purple circle classified as Seyfert and the outer $\sim 3~\mathrm{kpc}$ dominated by LINER emission. Line ratios in the purple circle are also plotted in the BPT diagrams, where the large error bars are due to the weak \Hb\ emission lines.}
\label{fig:fig5}
\end{figure*}

Based on our measurements of the entire galaxy datacube, we conducted a detailed analysis of the emission line properties. As illustrated in Figure~\ref{fig:fig4}, we present the narrow emission line flux and signal-to-noise ratio (SNR) maps for \Ha, \Hb, \OIII, \NII, \SII, and \OI\ following pPXF spectral fitting. The maps reveal a prominent extended ionized gas structure in the central bar region, while the spiral arms contain HII region clumps ionized by young stars. Notably, the \OIII\ emission is almost only detected in the EELR structure and does not extend to the spiral arms. The ionized region at the galaxy center exhibits a compact core component and is enclosed by a diffuse bi-conical structure. The SNR maps also display a similar distribution for \Ha, \NII, and \SII\ fluxes in the diffuse EELR region, whereas the \OIII\ flux appears to be enhanced in the galaxy's center.

Using the emission line measurements from our spectral fitting, we generated spatial distribution maps of line ratios log$_{10}$(\OIII/\Hb), log$_{10}$(\NII/\Ha), and log$_{10}$(\SII/\Ha) to investigate the ionization mechanisms across different regions of the galaxy. As shown in Figure~\ref{fig:fig5}, the top panels present the spatial distribution of log$_{10}$(\NII/\Ha) and log$_{10}$(\SII/\Ha) line ratios, revealing a clear distinction between the nuclear region and spiral arms. In the middle panel, we classify the ionization mechanisms for individual spaxels based on their BPT diagnostic positions, where different ionization sources are indicated by distinct colors. The EELR in the inner galaxy is classified as AGN in \NII/\Ha\ classification, while the spiral arms show ionization patterns characteristic of star formation. In the \SII/\Ha\ classification map, we observe a transition in the ionization mechanism, from Seyfert in the central regions to LINER characteristics towards the outer edges of the EELR structure. 
Spaxels classified as Seyfert dominate a $\sim 0.9~\mathrm{kpc}$ (3000 light-years) region in the galaxy center (inside the purple circle of the middle panel in Figure \ref{fig:fig5}). We also plot the line ratios inside the purple circle, where the error bars are large due to the weak \Hb\ emission lines. This extended structure is real for the FWHM of the seeing is $0.6^{\prime\prime}$ of the MUSE datacube, while this structure is $3^{\prime\prime}$, and it is consist with the enhanced \OIII\ region in the Figure \ref{fig:fig4}. This is an important evidence for recent agn activity in this galaxy. 

It is worth noting that the ionization mechanism of spectra classified as LINER remains controversial. In addition to ionization by low-luminosity AGNs \citep{Ho2008}, it may also originate from ultraviolet photons produced by evolved stellar populations (such as  post-AGB stars) \citep[e.g.][]{Yan2012,Singh2013,Belfiore2016} or shocks front passing through the ISM \citep{Heckman1980,Dopita2015}. TDEs with an elevated event rate have also been considered  as an alternative ionizing source of EELRs in TDE host galaxies \citep{Wevers2024a}. A recent study \citep{Mummery2025} shows that  ionizing radiation flux from a  TDE disk with a lifetime of $\mathcal{O}(10^2)$ yr may power a galactic scale EELR that lasts for $\mathcal{O}(10^4)$ yr, and contaminates a fraction of AGN identifications in BPT diagrams. Therefore, the ionization mechanism of the diffuse regions in the outer EELR warrants further modeling and analysis in the future. Specifically, studies combining spatially resolved stellar population types and luminosity measurements together with photoionization models could provide valuable insights.

\subsection{Gas and Stellar Kinematics} \label{subsec:sec3_3}

\begin{figure*}[ht!]
\begin{center}
\epsscale{1.0}
\plotone{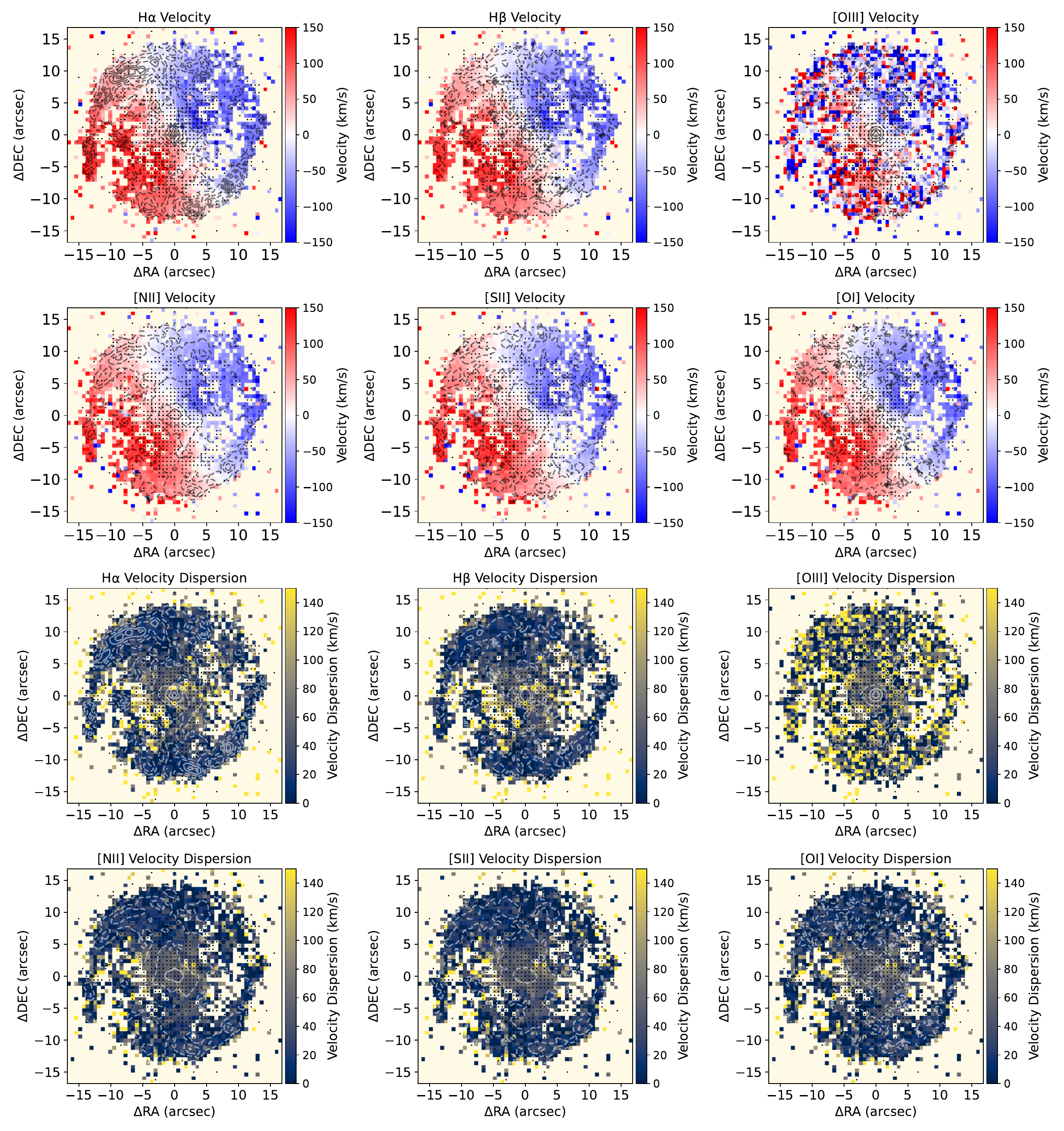}
\end{center}
\caption{The top six panels show the line-of-sight velocity distribution of the gas, while the bottom six panels display the gas velocity dispersion map. The contours represent the logarithmic flux distribution of different emission lines. Spaxels classified as AGN are marked with small black dots. Only regions with spectral SNR(\Ha) $>$ 5 are shown. Non-circular motions are observed in the EELR region, which is nearly parallel to the kinematic major axis, but there is also nearly zero line-of-sight velocity even outside the major axis. The velocity dispersion map shows higher gas dispersion within the EELR compared to the gas in the spral arms. Except in the \OIII velocity dispersion map, the dispersion values in the spiral arm are unreliable due to the low SNR.}
\label{fig:fig6}
\end{figure*}

We measured the line-of-sight velocity and velocity dispersion of gas and stellar components using pPXF fitting. The emission lines \Ha\ and \Hb\ are associated with one kinematic component, \NII, \SII\ and \OI\ with another, and \OIII\ as an independent component. The stellar continuum is treated as a separate kinematic component. The results are presented in Figure \ref{fig:fig6} for gas and Figure \ref{fig:figA4} for stellar components as a comparison. In the gas velocity map, we overlay different line flux contours. The map reveals that the EELR region is nearly parallel to the kinematic major axis, seems to have non-circular motion compared to the stellar velocity map. Within the EELR region, the gas kinematic major axis is distorted, with the gas velocity map showing nearly zero line-of-sight velocity even outside the major axis. Perpendicular to the major axis, an additional kinematic component of gas is observed compared to the stellar velocity map, with one side moving towards us and the other moving away. The gas velocity exceeds the stellar component by approximately $50\, \mathrm{km\,s^{-1}}$. The velocity dispersion map demonstrates lower gas dispersion within the EELR compared to the centrally peaked stellar dispersion. Decoupling of gas and stellar motions has been observed in the EELR hosts of the TDE iPTF–16fnl \citep{Wevers2024a} and QPEs \citep{Wevers2024b}, although our result is less pronounced. Several possibilities could explain these observations. First, the gas kinematics may be distorted by the galaxy bar potential \citep{Sellwood2010,Holmes2015,Feng2022}, while causing bar-induced inflow. Alternatively, the gas might originate from another galaxy through past minor merger events \citep{Barnes1996,Barrera-Ballesteros2015}, forming a misaligned gas disk (named ``inner warp''). In this scenario, the EELR gas resides above the galactic disk rather than within it, moving in a circular orbit with higher inclination than the stellar disk relative to the observer. This would cause the gas in the EELR region to primarily exhibit tangential velocity, resulting in nearly zero line-of-sight velocity, while being ionized by photons from the central AGN activity directed toward the observer. Additionally, since the observed emission lines are a brightness-weighted sum of ionized gas along the line of sight, geometric differences—such as a misalignment between the ionizing radiation cone and the stellar disk—can also affect the apparent velocity. It is difficult to observationally distinguish between an inner warp and radial inflow \citep{Zuleta2024}, further detailed kinematic modeling is necessary to substantiate these hypotheses.

\subsection{Ionizing Luminosity Estimation}\label{subsec:sec3_4}
\begin{figure*}[ht!]
\begin{center}
\epsscale{1.0}
\plotone{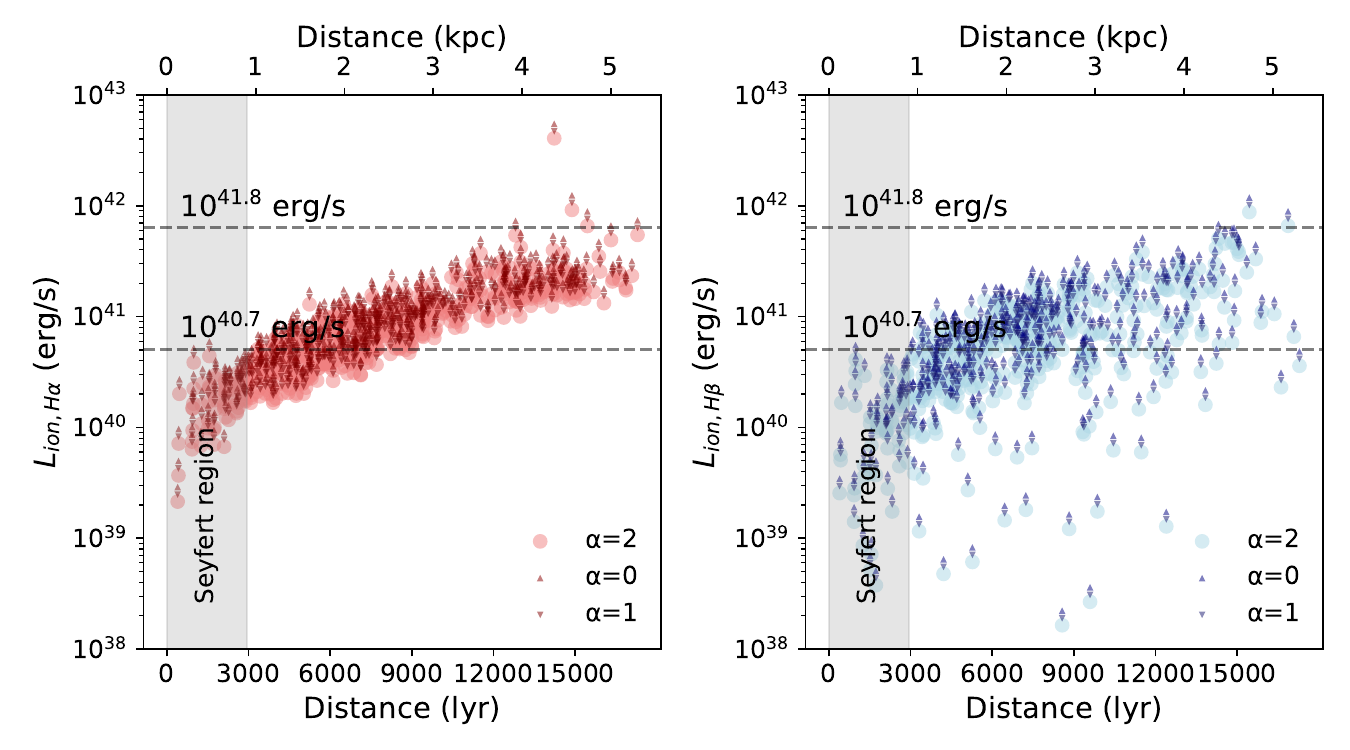}
\end{center}
\caption{Required ionizing luminosity inferred from \Ha\ (left) and \Hb\ (right) luminosities for each spaxel in the AGN-classified region as a function of projected physical distance from the central ionizing source. Ionizing spectra with different indices $\alpha$ are shown. The required ionizing luminosity is approximately $10^{41.8}\,\mathrm{erg\,s^{-1}}$ for all spaxels classified as AGN, assuming they are all ionized by a central point source. Spaxels in the central Seyfert region are marked with gray shading, indicating an ionizing luminosity of approximately $10^{40.7}\,\mathrm{erg\,s^{-1}}$.}
\label{fig:fig7}
\end{figure*}

Assuming that the EELR is ionized by the central AGN activity, we can estimate the required ionizing luminosity by relating it to the observed emission line luminosity. Similar methods were described in \citet{French2023} and \citet{Wevers2024b}, but we discribe a more detailed version in this paper. Here, we define the ionizing luminosity as:
\begin{equation}
L_{\text{ion}} = \int_{v_0}^{v_1} L_v \, dv
\end{equation}

where $L_{\nu}$\,($\mathrm{erg\,s^{-1}\,Hz^{-1}}$) represents the specific luminosity of the ionizing source at frequency $\nu$, $\nu_0$ and $\nu_1$ denotes the threshold frequency for hydrogen ionization, which we choose $\nu_0 = 13.6\,\mathrm{eV}/h$ and $\nu_1 = 54.4\, \mathrm{eV}/h$ (photons above this energy are assumed to ionize He\,{\sc ii}, h is Planck constant).

Based on photoionization models and the assumption of ionization-recombination equilibrium, we can establish a direct relationship between the ionizing luminosity and the observed emission line luminosity. Following \citet[Chapter 5, §5.10 \& Chapter 13, §13.5]{Osterbrock2006}, the total number of ionizing photons for hydrogen per second $Q(H^0)$ can be quantitatively linked to the \Hb\, luminosity $L_{\mathrm{H}\beta}$\,($\mathrm{erg\,s^{-1}}$) through

\begin{equation}
 Q(H^0) = \int_{\nu_0}^{\nu_1} \frac{L_{\nu}}{h \nu} d\nu =
{ \left(h \nu_{\mathrm{H}\beta} \frac{\alpha_{\mathrm{H}\beta}^{\mathrm{eff}}(\mathrm{H}^0, T)}{\alpha_B(\mathrm{H}^0, T)} \frac{\Omega}{4\pi}\right)}^{-1}L_{\mathrm{H}\beta}\ ,
\end{equation}
where $h\nu$ represents the energy of ionizing photons, $h\nu_{\mathrm{H}\beta}$ is the energy of H$\beta$ photons. The effective recombination coefficient for H$\beta$ emission is given by $\alpha_{\mathrm{H}\beta}^{\mathrm{eff}}(\mathrm{H}^0, T)$, while $\alpha_B(\mathrm{H}^0, T)$ represents the total recombination coefficient for hydrogen. Assuming Case B recombination at $T = 10^4\,\mathrm{K}$, we have $\alpha_{\mathrm{H}\beta}^{\mathrm{eff}} = 3.03 \times 10^{-14}\,\mathrm{cm^3\,s^{-1}}$ and $\alpha_B = 2.59 \times 10^{-13}\,\mathrm{cm^3\,s^{-1}}$\citep{Osterbrock2006}. In such case, we can also associate \Ha\, luminosity with \Hb\, luminosity using the ratio $L_{\mathrm{H}\alpha}/L_{\mathrm{H}\beta} = 2.87$, allowing us to estimate ionizing luminosity though $L_{\mathrm{H}\alpha}$ too.
The factor $\frac{\Omega}{4\pi}$ represents the covering fraction of each spaxel in the MUSE data cube, which can be understood as the opening solid angle of the gas cloud within each spaxel relative to the central ionizing source. We follow the definition used in \citet{French2023}

\begin{equation}
f = \frac{\Omega}{4\pi} = \frac{(2 \arctan(0.5/r))^2}{4\pi} \ .
\end{equation}

Assuming the ionizing continuum follows a power-law distribution $L_{\nu} = C\nu^{-\alpha}$, the relationship between $L_{\text{ion}}$ and $Q(H^0)$ can be derived as
\begin{equation}
L_{\text{ion}} = 
\begin{cases} 
Q(H^0) \frac{h\alpha}{\alpha - 1} \cdot \frac{v_0^{1 - \alpha} - v_1^{1 - \alpha}}{v_0^{-\alpha} - v_1^{-\alpha}} & (\alpha \neq 0, 1) \\[2ex]
Q(H^0) \frac{h(v_1 - v_0)}{\ln\left(v_1/v_0 \right)} & (\alpha = 0) \\[2ex]
Q(H^0) \frac{h\ln\left(v_1/v_0\right)}{v_0^{-1} - v_1^{-1}} & (\alpha = 1)
\end{cases}
\end{equation}
The detailed derivation of this relationship is presented in Appendix \ref{appendix:A}.

Using the aforementioned equations, we estimate the required ionizing luminosities $L_{\text{ion},\mathrm{H}\alpha}$ and $L_{\text{ion},\mathrm{H}\beta}$ for all spaxels classified as AGN in the \NII/\Ha\ BPT diagram, based on the measured \Ha\ and \Hb\ luminosities. The results are presented in Fig \ref{fig:fig7}. We adopt three typical AGN EUV spectral indices $\alpha$ \citep{Stevans2014} to calculate the ionizing luminosities and demonstrated that the choice of $\alpha$ does not affect the conclusion. The required ionizing luminosity increases with distance from the center, with an ionizing luminosity of approximately $L_{\text{ion}} =10^{41.8}\,\mathrm{erg\,s^{-1}}$ being sufficient to ionize all spaxels classified as AGN, and a lower ioninzing luminosity about $L_{\text{ion}}=10^{40.7}\,\mathrm{erg\,s^{-1}}$ is needed in the center $0.9~\mathrm{kpc}$ Seyfert region. It is important to note that due to the low signal-to-noise ratio of individual \Hb\ spectra, we did not perform spaxel-by-spaxel attenuation correction for the emission line fluxes. Therefore, this ionizing luminosity represents only a lower limit. However, by summing the fluxes of all spaxels in the AGN region and calculating the \Ha/\Hb\ ratio, we estimated an overall attenuation value of approximately $A_{\rm V}^{\text{gas}} = 0.15$, which does not significantly affect the estimation of the ionizing luminosity. 

Due to the finite speed of light, gas at different distances from the center is ionized by photons emitted during different epochs of AGN activity. Therefore, the EELR can reflect the history of AGN activity, or its duty cycle. By comparing the current AGN ionizing luminosity with the ionizing luminosity estimated from emission lines, we can preliminarily determine whether the emission lines can be ionized by the current AGN activity. Given that EUV is heavily absorbed by the interstellar medium, we use X-ray luminosity for an approximate estimation. \citet{Nicholl2020} observed with the Swift X-ray Telescope (XRT) and reported that the $0.3 - 10\, \mathrm{keV}$ luminosity in early 2019 was $L_{\text{X}} = 5.1 \times 10^{40}\,\mathrm{erg\,s^{-1}}$. In \citet{Nicholl2024}, higher resolution Chandra observations revealed an additional X-ray source located $7^{\prime\prime}$ southeast of the host galaxy nucleus. Excluding its influence, the quiescent X-ray luminosity of the nuclear region is $L_{\text{X}} \leq 7.2 \times 10^{39}\,\mathrm{erg\,s^{-1}}$. X-ray luminosity and UV ionizing luminosity are in different bands, therefore considering a bolometric correction factor of approximately 10 \citep{Duras2020}, the observed current quiescent bolometric luminosity $L_{\text{bol}} \leq  10^{40.8}\,\mathrm{erg\,s^{-1}}$ is still lower than the ionizing luminosity inferred from the emission lines within the EELR ($10^{41.8}\,\mathrm{erg\,s^{-1}}$), but close to the ionizing luminosity estimated for the Seyfert region ($10^{40.7}\,\mathrm{erg\,s^{-1}}$). This suggests that the diffuse emission line region within the EELR cannot only be explained by the current nuclear activity, possibly indicating a decline in recent AGN activity, like recently faded AGN. However, the central $\sim 0.9~\mathrm{kpc}$ region may be ionized by the remaining weak AGN activity, corresponding to a timescale of approximately 3000 years.

\section{Summary} \label{sec:sec4}
In this study, we present MUSE integral-field spectroscopy data analysis of an EELR in the host galaxy of TDE AT2019qiz followed by QPEs. The EELR extends to approximately 3.7 kpc, with a central compact region dominated by \OIII\ emission and a diffuse \NII\ emission region in the outskirts.

We performed continuum and emission line fitting on the integrated spectra of the entire EELR and three localized regions within it, as well as the galactic center with broad line components from previous TDE. Our analysis reveals that the outer regions of the EELR are classified as LINER in BPT diagrams, while the central region exhibits Seyfert characteristics. Spaxel-by-spaxel spectral measurements confirm this spatial distribution, with the central 0.9 kpc dominated by Seyfert-like emission and the outer regions showing LINER-like properties.

Stellar population analysis indicates that the galaxy experienced a star formation episode approximately 1 Gyr ago, after which star formation quenched, displaying typical post-starburst galaxy characteristics. Stellar and gas kinematics measurements reveal non-circular gas velocity fields with additional components compared to the stellar velocity field, potentially resulting from the influence of a bar potential or past minor merger activity.

Assuming the EELR is ionized by a central AGN and based on photoionization-recombination equilibrium, we estimated the required ionizing luminosity using \Ha\ and \Hb\ emission lines. Our results indicate a lower limit of $10^{41.8}\,\mathrm{erg\,s^{-1}}$ to ionize all spaxels within the EELR, while ionizing just the Seyfert region requires at least $10^{40.7}\,\mathrm{erg\,s^{-1}}$. The currently observed nuclear luminosity is insufficient to produce such strong ionization, suggesting past AGN activity that has since diminished, similar to a recently faded AGN scenario. 

It should be noted that AGN as an ionizing source is merely an assumption here. Alternative sources such as past TDEs or evolved stars cannot be ruled out (as we discussed in Section~\ref{subsec:sec3_2}), and the EELR may also result from a combination of multiple ionizing mechanisms. The primary aim of this study is to demonstrate the existence of the EELR in this galaxy, while additional observational evidence and more detailed modeling are needed to definitively determine the ionizing source. Nevertheless, considering the presence of mid-infrared echoes in the WISE light curve suggesting a dusty torus \citep{Short2023,Pasham2025}, the assumption of past AGN activity is reasonable.

In TDE-QPE host galaxies, additional evidence for recent AGN activities has been inferred from QPE timing in the framework of EMRI+disk model \citep{Zhou2024a,Zhou2024b,Zhou2025}. As in the unified scenario proposed by \citet{Jiang2025}, AGN activities may boost both the TDE rate and the formation rate of low-eccentricity EMRIs, consequently boost the occurrence of QPEs. Our finding of an EELR in the first clear TDE-QPE associated galaxy provides strong support for the unified model. In the near future, this scenario can be verified by more advanced IFS observations of the host galaxies of TDE-QPE associations, which are expected to be unveiled by various surveys in the golden era of time-domain astronomy. On the other hand, the detection of more distant EELRs requires excellent IFS capabilities with sufficient spatial resolution. Thus at present, only instruments like VLT/MUSE can effectively resolve these extended structures. Additionally, EELRs represent ideal targets for the upcoming China Space Station Telescope (CSST) \citep{Zhan2011,CSSTCollaboration2025} integral-field spectrograph (CSST-IFS), which features a $6^{\prime\prime} \times 6^{\prime\prime}$ field of view, high spatial resolution of $0.2^{\prime\prime}$ $\textrm{spaxel}^{-1}$, and wavelength coverage of 0.35-1.0 $\micron$. This instrument will facilitate EELR observations of more TDE and QPE samples, enabling detailed emission line, stellar population, and kinematic measurements.

\begin{acknowledgments}
We thank the referee for very positive and constructive
comments, which have improved the manuscript significantly.
We thank Dr. Zhi Li and Dr. Lin Lin for their discussions and suggestions on galactic dynamics and barred galaxies. We thank Renhao Ye for providing useful data analysis advice. This work is supported by National Key R\&D Program of China No.2022YFF0503402, and the National Natural Science Foundation of China (NSFC, Nos. 12233005), the Youth Innovation Promotion Association CAS (id. 2022260). N.J. acknowledges the support by the National Natural Science Foundation of China (grants 12192221,12393814), the Strategic Priority Research Program of the Chinese Academy of Sciences (XDB0550200). 
Based on data obtained from the ESO Science Archive Facility under request number 957350.
\facility {ESO/VLT:MUSE}
\software{Astropy, Matplotlib} \citep{astropyCollaboration2013, astropyCollaboration2018,Hunter2007}
\end{acknowledgments}

\appendix
\counterwithin{figure}{section}
\section{Derivation of the Relationship between Ionizing Luminosity and Ionizing Photon Rate for a Power-Law Spectrum} \label{appendix:A}

In this section, we derive the relationship between the ionizing luminosity and the number of ionizing photons per second (ionizing photon rate), assuming the ionizing spectrum follows a power-law distribution of the form $L_{\nu} = C\nu^{-\alpha}$, where $C$ is a normalization constant and $\alpha$ is the power-law index.

\paragraph{Case 1: $\alpha \neq 0, 1$}
The ionization luminosity $L_{\text{ion}}$ is obtained by integrating the specific luminosity $L_v$ over the frequency range from $v_0$ to $v_1$:
\[
L_{\text{ion}} = \int_{v_0}^{v_1} L_v \, dv = \int_{v_0}^{v_1} C v^{-\alpha}\,dv = \frac{C}{\alpha-1}\Big(v_0^{-(\alpha-1)}-v_1^{-(\alpha-1)}\Big)
\]

The photon rate $Q$ is calculated by integrating the specific luminosity over frequency and dividing by the photon energy:
\[
Q = \int_{v_0}^{v_1} \frac{L_v}{h v} \, dv 
= \int_{v_0}^{v_1} \frac{C\, v^{-\alpha}}{h\, v} \, dv 
= \frac{C}{h} \int_{v_0}^{v_1} v^{-(\alpha + 1)} \, dv 
= \frac{C}{h} \cdot \frac{v_0^{-\alpha} - v_1^{-\alpha}}{\alpha}
\]

Solving for constant $C$:
\[
C = \frac{Q h \alpha}{v_0^{-\alpha} - v_1^{-\alpha}}
\]

Substituting $C$ into the expression for $L_{\text{ion}}$:
\[
L_{\text{ion}} = \frac{Q h\alpha}{\alpha - 1} \cdot \frac{v_0^{1 - \alpha} - v_1^{1 - \alpha}}{v_0^{-\alpha} - v_1^{-\alpha}}
\]

\paragraph{Case 2: $\alpha = 0$}
The specific luminosity becomes constant:
\[
L_v = C v^{0} = C
\]

The ionization luminosity:
\[
L_{\text{ion}} = \int_{v_0}^{v_1} C \, dv = C (v_1 - v_0)
\]

The photon rate:
\[
Q = \int_{v_0}^{v_1} \frac{C}{h v} \, dv = \frac{C}{h} \ln\left( \frac{v_1}{v_0} \right)
\]

Solving for constant $C$ and substituting it into the expression for $L_{\text{ion}}$:
\[
L_{\text{ion}} = \frac{Q h (v_1 - v_0)}{\ln\left(v_1/v_0\right)}
\]

\paragraph{Case 3: $\alpha = 1$}
The specific luminosity is:
\[
L_v = C v^{-1}
\]

The ionization luminosity:
\[
L_{\text{ion}} = \int_{v_0}^{v_1} C v^{-1} \, dv = C \ln\left( \frac{v_1}{v_0} \right)
\]

The photon rate:
\[
Q = \int_{v_0}^{v_1} \frac{Cv^{-1}}{h v} \, dv = \frac{C}{h} \left( \frac{1}{v_0} - \frac{1}{v_1} \right)
\]

Therefore:
\[
L_{\text{ion}} =  \frac{Qh\ln\left(v_1/v_0\right)}{v_0^{-1} - v_1^{-1}}
\]

In summary, the relationship between ionization luminosity $L_{\text{ion}}$ and photon flux $Q$ can be expressed as a piecewise function:

\[
L_{\text{ion}} = 
\begin{cases} 
Q \frac{h\alpha}{\alpha - 1} \cdot \frac{v_0^{1 - \alpha} - v_1^{1 - \alpha}}{v_0^{-\alpha} - v_1^{-\alpha}} & (\alpha \neq 0, 1)  \\[2ex]
Q \frac{h(v_1 - v_0)}{\ln\left( v_1/v_0 \right)} & (\alpha = 0) \\[2ex]
Q \frac{h\ln\left( v_1/v_0\right)}{v_0^{-1} - v_1^{-1}}  & (\alpha = 1)
\end{cases}
\]

\begin{figure}[ht!]
\begin{center}
\epsscale{1.0}
\plotone{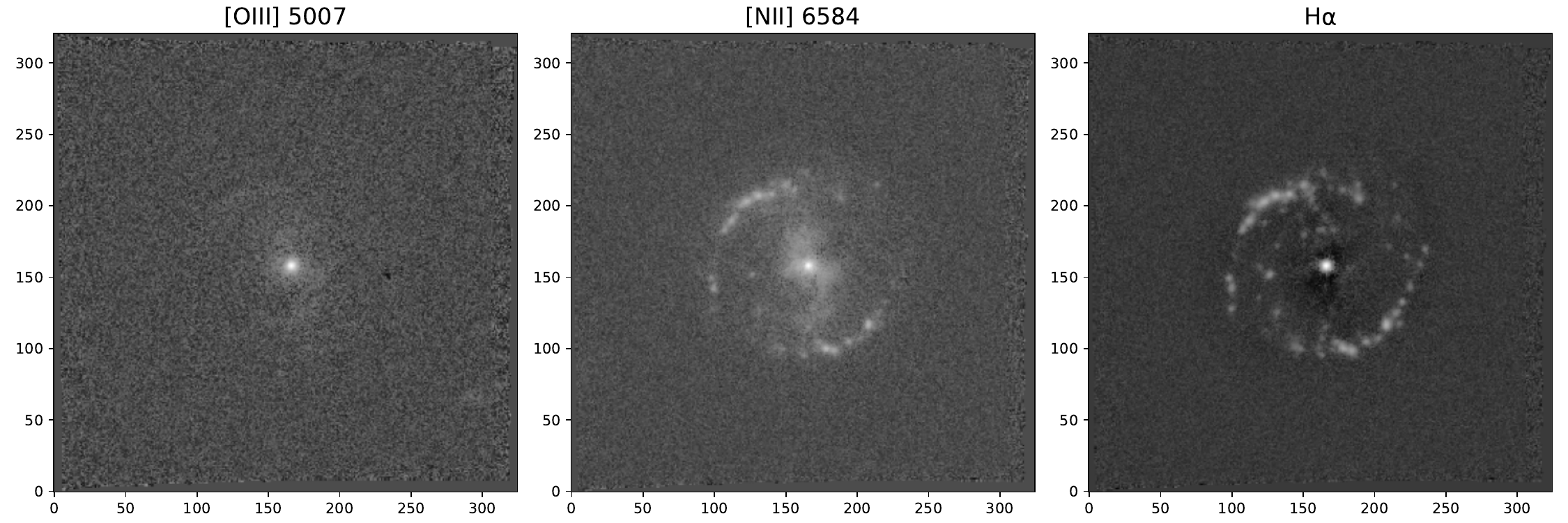}
\end{center}
\caption{Maps of \OIII $\lambda$5007, \Ha, and \NII $\lambda$6584 at the original resolution of the MUSE datacube. These maps are derived by integrating the emission line regions and subtracting the adjacent continuum, thus not accounting for stellar absorption but avoiding potential artifacts from poor Gaussian fitting. A prominent \NII-dominated emission line region is evident at the galaxy center, also present in the \OIII\ map but weaker. In the \Ha\ map,  there are regions appear black due to stellar absorption lines. After subtracting the stellar continuum, Gaussian fitting reveals the \Ha\ component of the EELR in Figure \ref{fig:fig4}.}
\label{fig:figA1}
\end{figure}

\begin{figure*}[ht!]
\begin{center}
\epsscale{0.8}
\begin{tabular}{cc}
\includegraphics[width=0.45\textwidth]{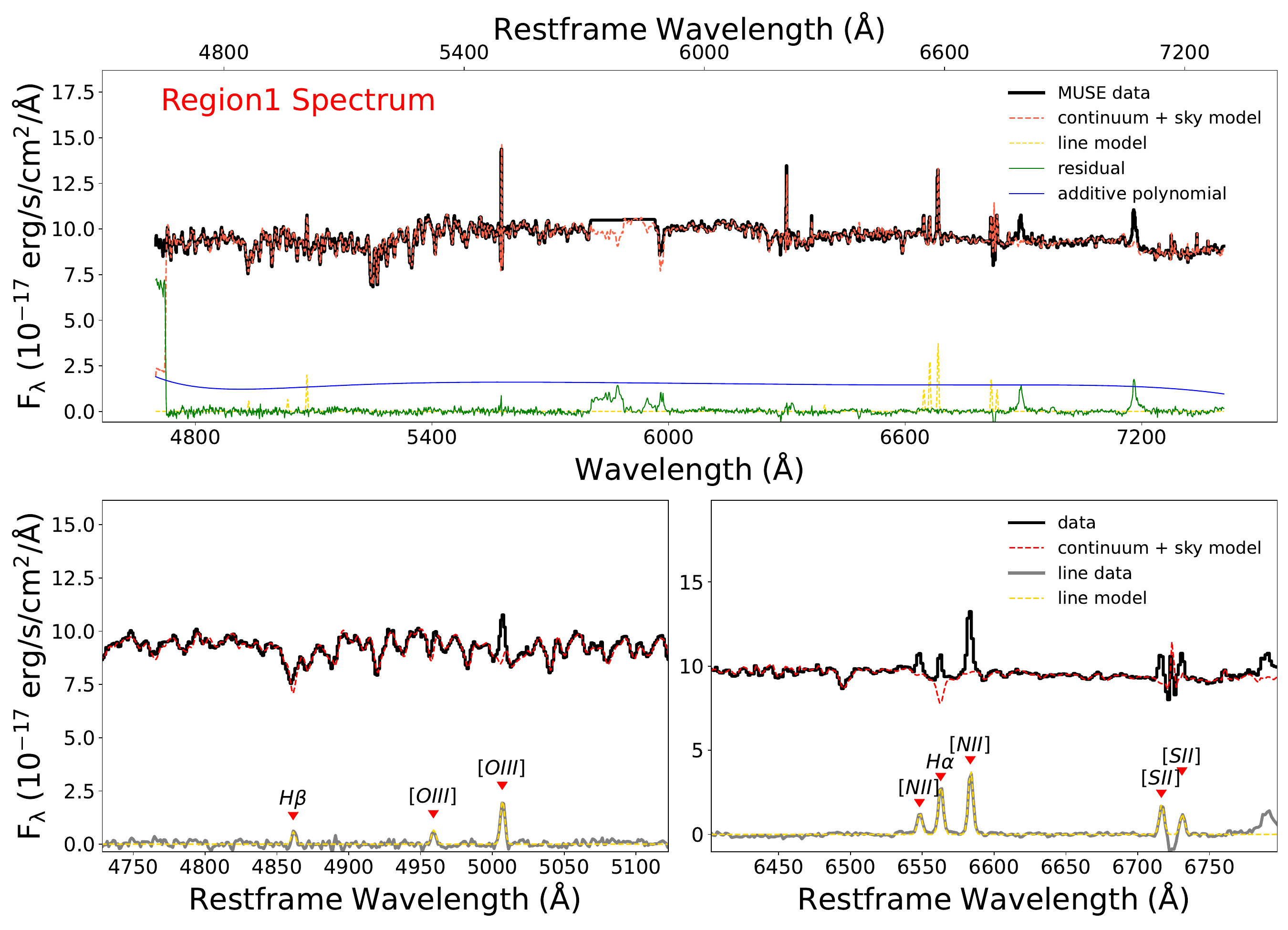} &
\includegraphics[width=0.45\textwidth]{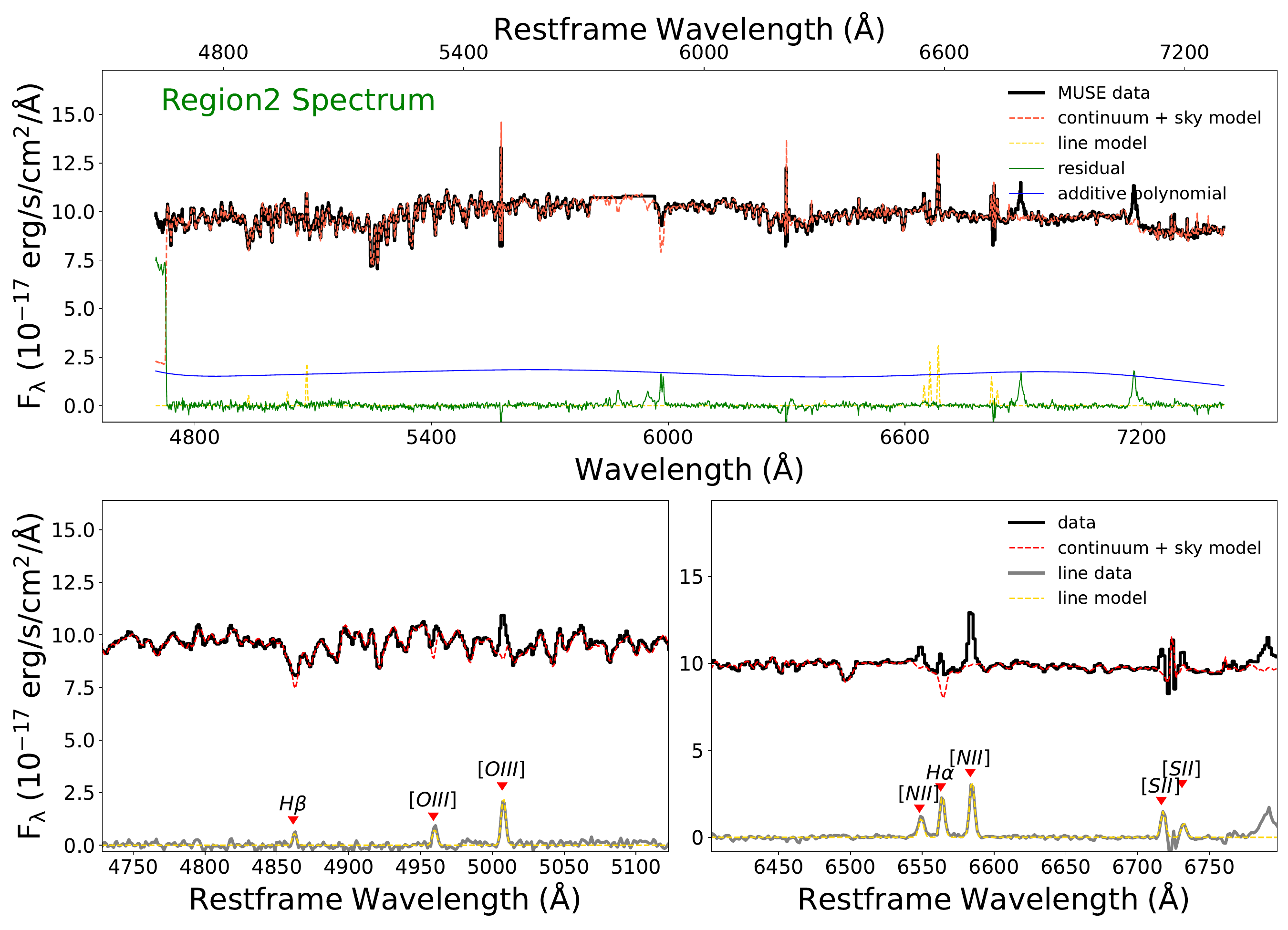} \\
\includegraphics[width=0.45\textwidth]{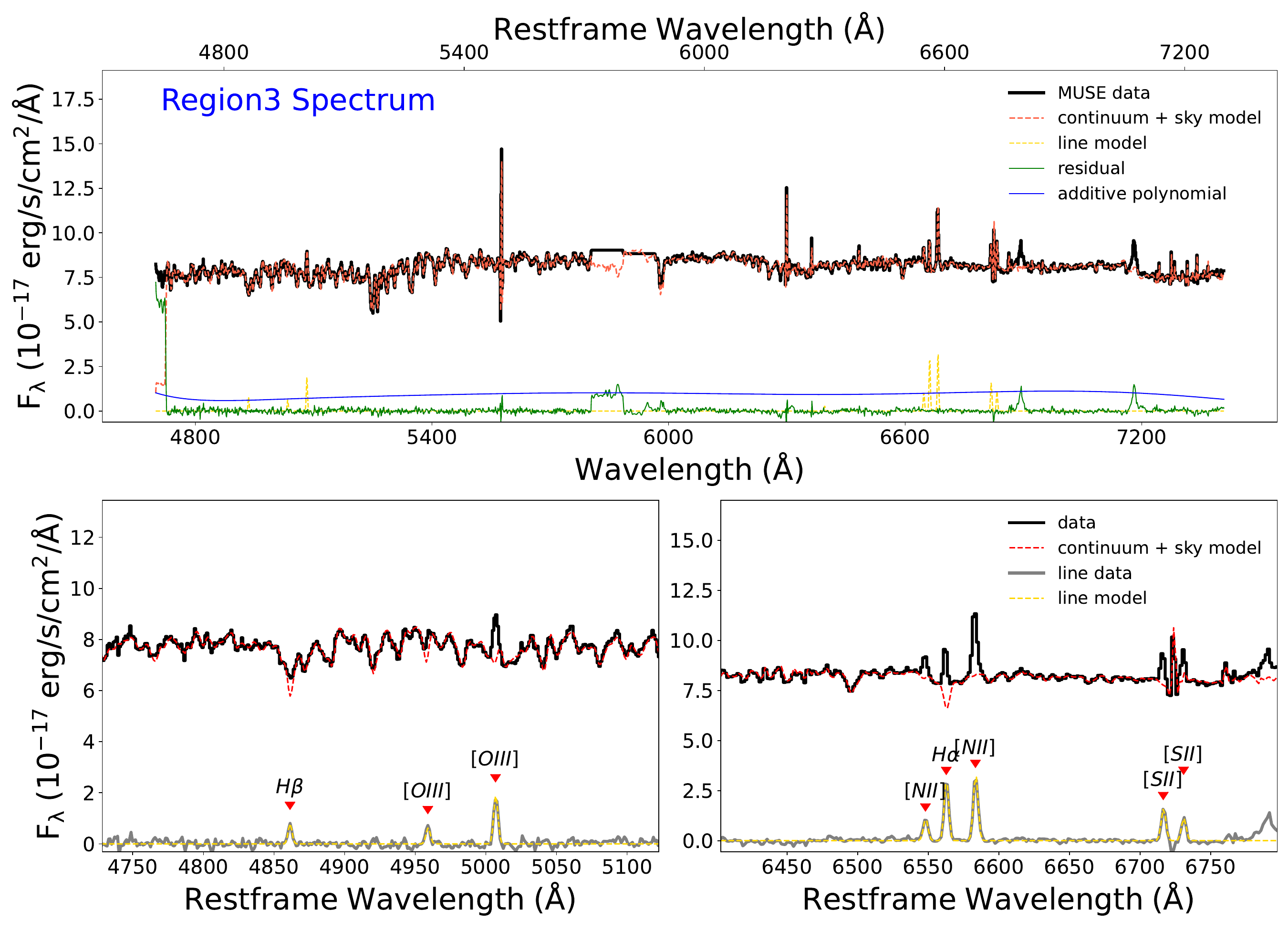} &
\includegraphics[width=0.45\textwidth]{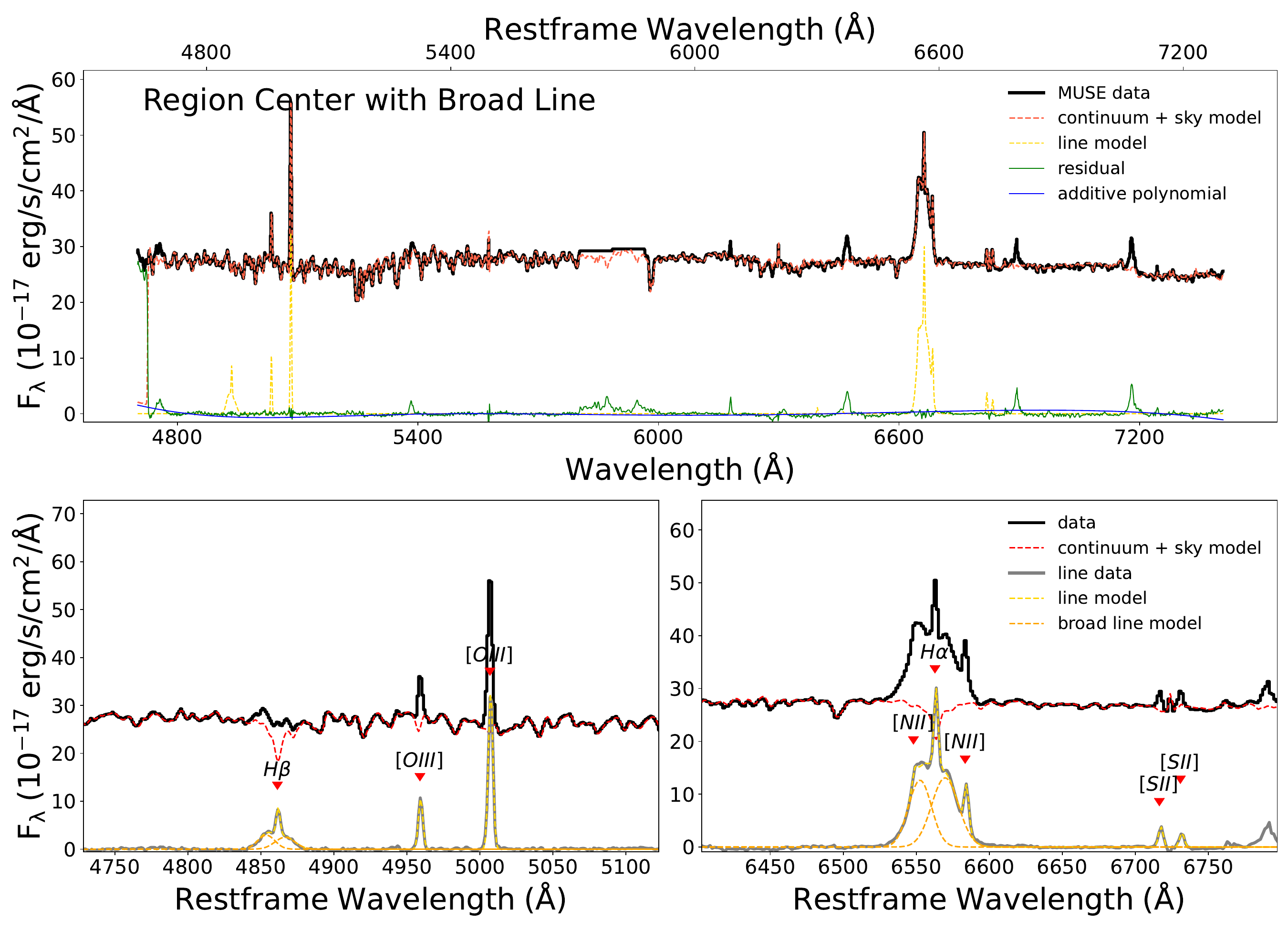}
\end{tabular}
\end{center}
\caption{Spectral fitting results for different regions of the extended emission line region. The panels show Region 1 (top left), Region 2 (top right), Region 3 (bottom left), and the central region with broad line (bottom right). In each panel, the black solid line represents the MUSE spectral data, the red dashed line shows the continuum+sky model, the gray solid line indicates the emission line data (obtained by subtracting the continuum+sky model from the data), and an 8-order additive polynomial is shown in blue solid line. The yellow dashed line represents the emission line model, the orange dashed line shows the broad line model, and the green line displays the fitting residuals. }
\label{fig:figA2}
\end{figure*}

\begin{figure}[ht!]
\begin{center}
\epsscale{1.0}
\plotone{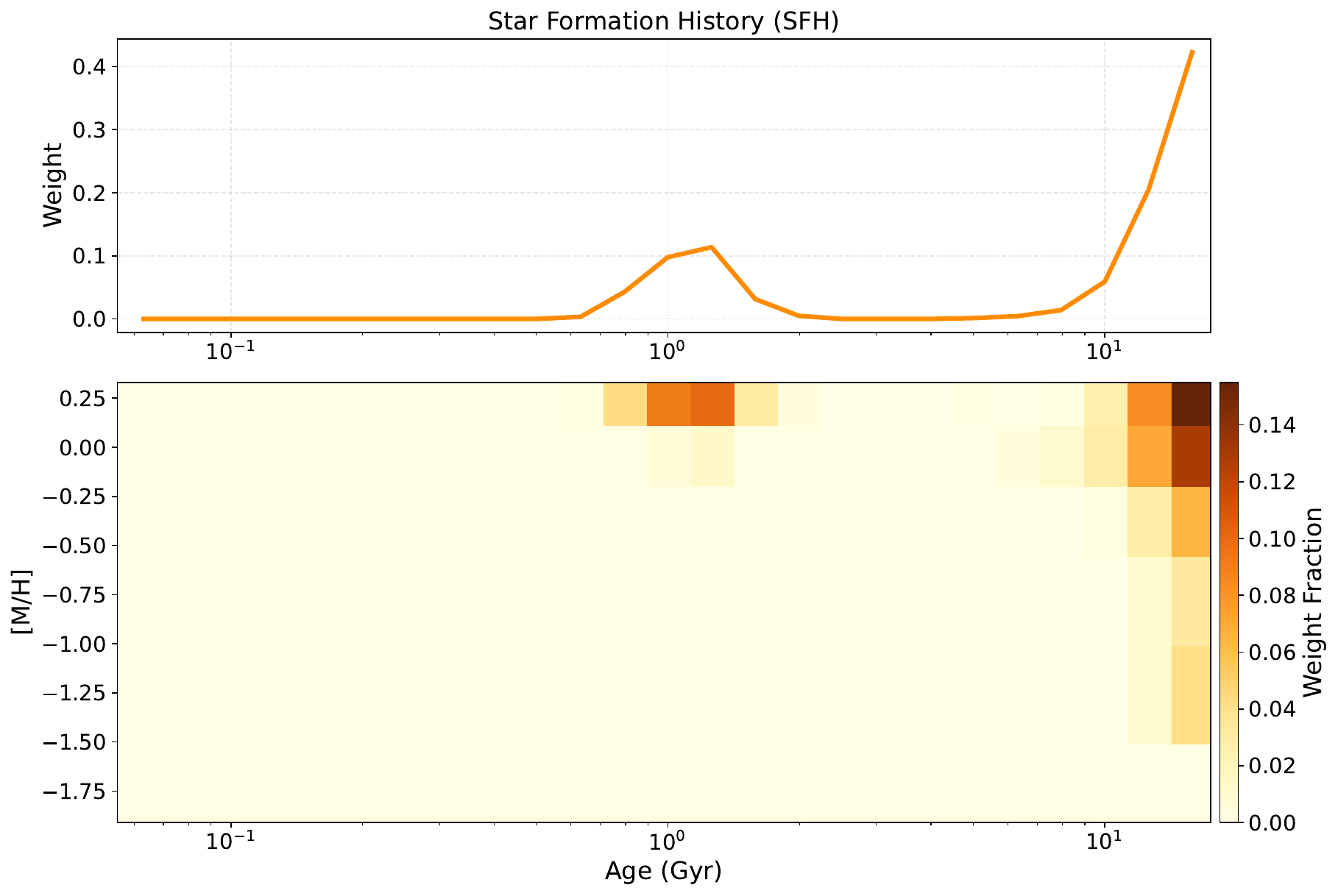}
\end{center}
\caption{Stellar population analysis using pPXF fitting for the entire EELR region defined between the solid and dashed white circles in Figure \ref{fig:fig1}. The bottom panel displays the weight map of E-MILES stellar population spectral templates across different ages and metallicity, where $[\mathrm{M}/\mathrm{H}]$ is the stellar metallicity relative to the solar value, defined as $[\mathrm{M}/\mathrm{H}] = \log_{10}\left( N_\mathrm{M}/N_\mathrm{H} \right)_{\text{star}} - \log_{10}\left( N_\mathrm{M}/N_\mathrm{H} \right)_{\odot}$, where $N$ is the number density of atoms. The panel above shows the weight distribution as a function of age (star formation history), with metallicity weights integrated within each age bin. A prominent star formation event approximately 1 Gyr ago is evident, characteristic of a typical post-starburst galaxy.}
\label{fig:figA3}
\end{figure}

\begin{figure}[ht!]
\begin{center}
\epsscale{1.0}
\plotone{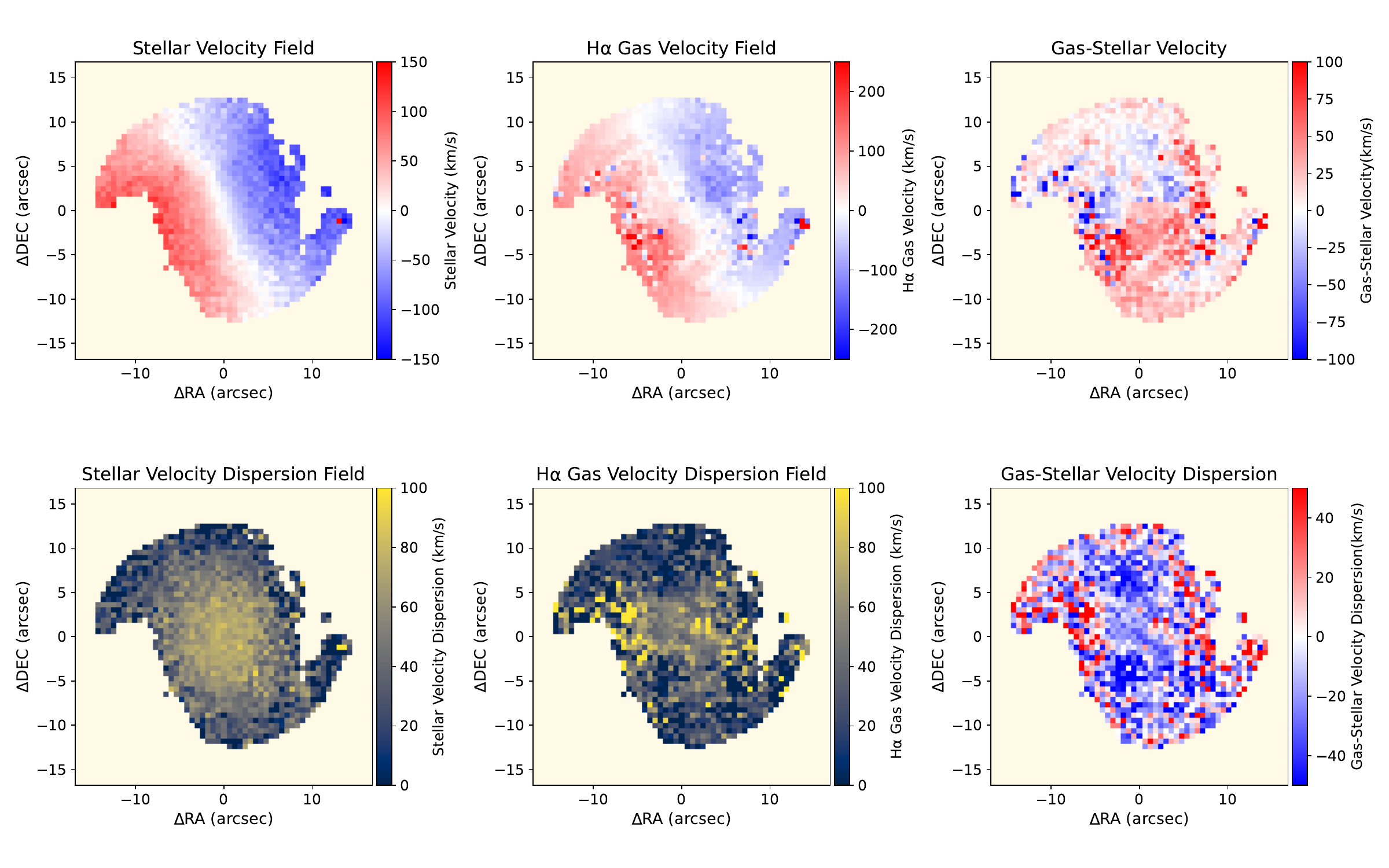}
\end{center}
\caption{Comparison between the stellar and \Ha\ gas kinematics. Left: stellar velocity and velocity dispersion. Middle: \Ha\ gas velocity and velocity dispersion. Right: Gas velocity (dispersion) minus stellar velocity (dispersion). The line-of-sight velocity shows slight differences for gas, with noncircular motion inside the EELR, and an additional kinematic component observed perpendicular to the major axis, with the right side exhibiting blueshift and the left side exhibiting redshift. The velocity dispersion map demonstrates lower gas dispersion within the EELR compared to the centrally peaked stellar dispersion. Only regions with summed spectral SNR $>$ 10 are shown.}
\label{fig:figA4}
\end{figure}

\bibliography{reference}{}
\bibliographystyle{aasjournal}
\end{CJK*}
\end{document}